\documentclass[aps,prl,twocolumn,superscriptaddress,floatfix]{revtex4-1}

\usepackage{graphicx}% Include figure files
\usepackage{dcolumn}% Align table columns on decimal point
\usepackage{bm}% bold math
\usepackage{xcolor}

\begin{document}

%\title{Resonant-tunneling activated shallow impurities in GaAs quantum wells and their single-photon emission}% Force line breaks with \\
\title{Tunneling blockade and single-photon emission in GaAs double quantum wells}% Force line breaks with \\

\author{M. Yuan} 
\email{yuan@pdi-berlin.de}
\author{A. Hern{\'a}ndez-M{\'i}nguez}
\author{K. Biermann}
\author{P. V. Santos}
\affiliation{%
 Paul-Drude-Institut f{\"u}r Festk{\"o}rperelektronik, 
Leibniz-Institut im Forschungsverbund Berlin e. V., 
Hausvogteiplatz 5-7, 10117 Berlin, Germany
}%

\date{\today}% It is always \today, today,

\begin{abstract}
We report on the selective excitation of single impurity-bound exciton states in a GaAs double quantum well (DQW). The structure consists of two quantum wells (QWs) coupled by  a thin tunnel barrier. The DQW  is subject to a transverse electric field to create spatially indirect inter-QW excitons with electrons and holes located in different QWs. 
We show that the presence of intra-QW charged excitons (trions) blocks carrier tunneling across the barrier to form indirect excitons, thus opening a gap in their  emission spectrum. This behavior is attributed to the low binding energy of the trions.
Within the tunneling blockade regime, emission becomes dominated by processes involving excitons bound to single shallow impurities, which behave as two-level centers  activated by resonant tunneling.  The quantum  nature of the emission is confirmed by the anti-bunched photon emission statistics. The narrow distribution of emission energies ($\sim 10$~meV) and the electrical connection to the QWs make these single-exciton centers interesting candidates for applications in single-photon sources. 

\end{abstract}

\pacs{Valid PACS appear here}% PACS, the Physics and Astronomy
\maketitle

%%%%%%%%%%%%%%%%%%%%%%%%%%%%%%%%%%%%%%%%%%%%%%%%%%%%%%%%
%\section{Introduction}
The interplay between resonant tunneling and inter-particle interactions in low-dimensional quantum systems gives rise to interesting phenomena in the transport of single particles. A prototype example is the Coulomb blockade, where repulsive Coulomb interaction blocks the transport through a localized state between two reservoirs. \cite{Zeller_PR_181_789_69,Kastner_RMP_64_859_92}  This blockade, which expresses itself as plateaus between steps in the current {\it versus} voltage characteristics, is a direct signature of single carrier transport.

\begin{figure}
\includegraphics[width=0.9\columnwidth, angle=0, clip]{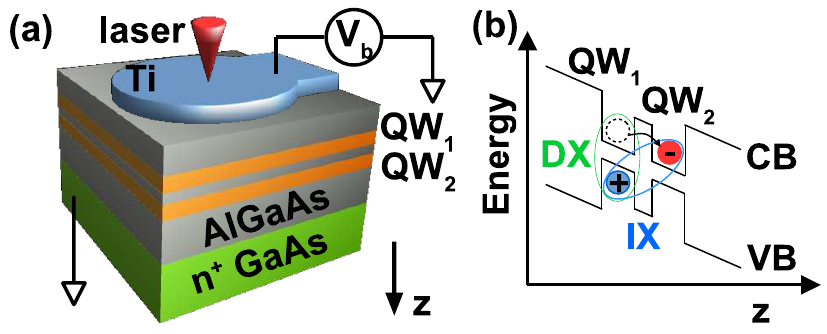}
\caption{(a) (Al,Ga)As DQW structure on doped GaAs (001) substrate. The DQW is embedded in the intrinsic region of a  Schottky diode defined by the  semitransparent Ti top contact. The bias voltage $V_b$ is applied between the top contact and the doped substrate.  The photoluminescence (PL) studies were carried out at 4.2~K by exciting carriers using a focused laser spot. 
(b) Energy band diagram of the DQW showing the conduction (CB) and valence bands (VB) profiles under an electric field, which leads to IX formation.
}
\label{setup}
\end{figure}

In this paper,  we report on an analogous blockade phenomenon during the transport of exciton-related species across a thin tunnel barrier between two semiconductor quantum wells (QWs)   (cf. Fig.~\ref{setup}(a)).  Each QW supports intra-QW neutral excitons (denoted here as direct excitons, DX) as well as intra-QW charged excitons (trions, T).  A transverse electric field $F_z$ applied across the double QW (DQW) structure controls the tunneling probability and enables the creation of spatially indirect, inter-QW excitons (IXs) consisting of electrons and holes resident in different QWs (cf.~Fig.~\ref{setup}(b)). The energy of IX is tunable by $F_z$ due to quantum-confined Stark effect (QCSE). \cite{Miller_PRL53_1984,Chen_PRB36_1987}
We show that the low binding energy of trions blocks carrier tunneling over a range of applied fields and,  thus, the formation of IX states. 
Unlike Coulomb blockade, which arises from the repulsive interaction between electrons, the tunneling blockade reported here originates from the attractive exciton-carrier interaction leading to trion formation. 
Within the blockade regime, tunneling becomes restricted to resonant processes leading to the excitation of individual excitons bound to shallow impurities. 
These states act as electrically controlled two levels systems: their  quantum nature is evidenced by the anti-bunched photon emission statistics. 
The electric control and narrow distribution of emission energies make these bound-exciton emitters potential candidates for GaAs-based single-photon sources.

%%%%%%%%%%%%%%%%%%%%%%%%%%%%%%%
% Figure 1: setup
%%%%%%%%%%%%%%%%%%%%%%%%%%%%%%%

% ------------------------------------------------------------------------------
% experimental details

The studies were carried out on the (Al,Ga)As structure illustrated in Fig.~\ref{setup}(a). The sample was grown by molecular beam epitaxy (MBE) on n-doped GaAs(001) substrate.  The DQW consists of two $16$~nm-wide GaAs QWs separated by a 4~nm-thick Al$_{0.33}$Ga$_{0.67}$As barrier. The DQW is subjected to an electric field $F_z$ generated by a bias voltage  $V_b$ applied between a 10~nm-thick semi-transparent top titanium contact\cite{Huber_PSS166_1998,Rapaport_PRB72_075428_05,Hammack_JAP99_66104_06} and the n-doped substrate. 

\begin{figure}
\includegraphics*[width=0.9\columnwidth, angle=0, clip]{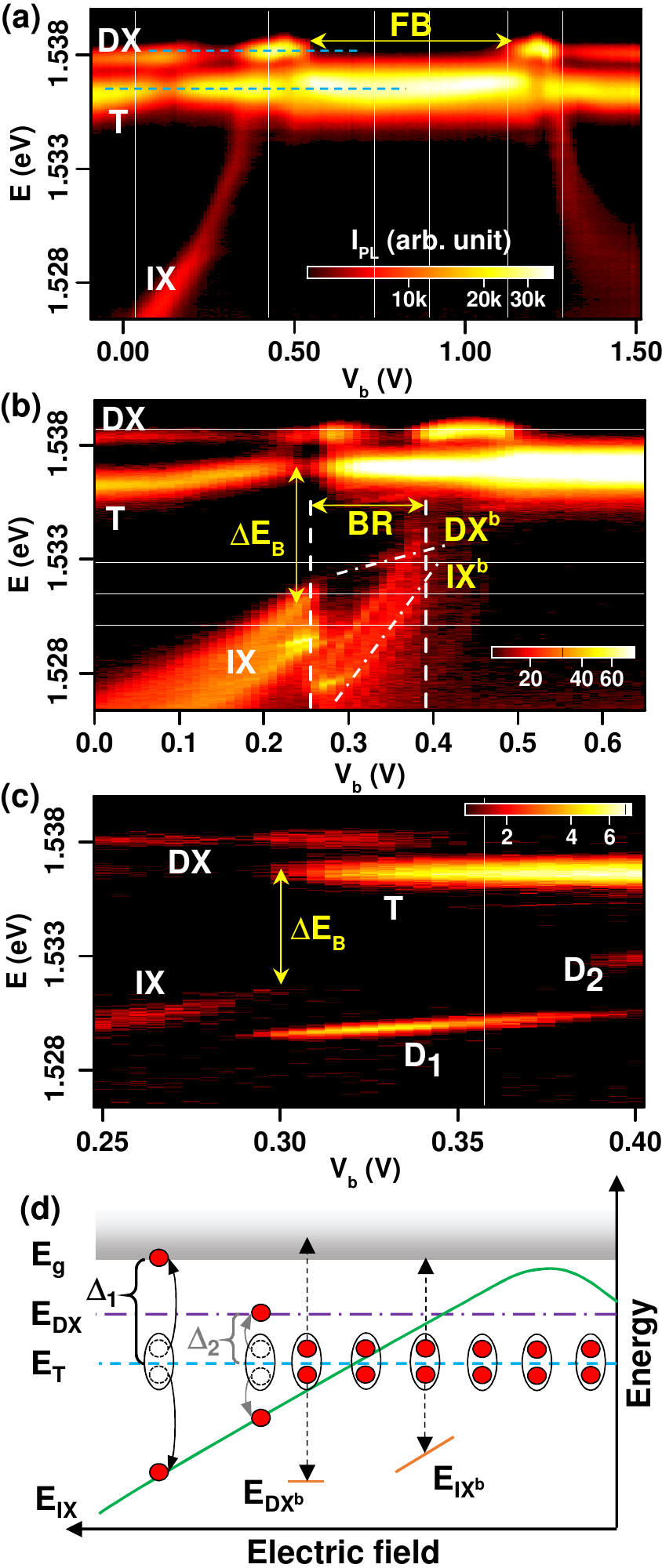}
\caption{\label{PLV}(a) Spatially integrated PL vs. bias voltage $V_b$ recorded under a high laser power ($P=240$~ nW). FB denotes the flat band region.
(b) Idem, but recorded under a low power ($P=0.6$~nW), for bias in the IX-FB transition region of (a). The formation of IX is blocked within the spectral windows indicated by $\Delta E_B$. Individual lines DX$^b$ and IX$^b$ associated with excitons bound to shallow impurities appear in the blockade region BR between the FB and IX regimes. (c) Bias maps recorded with high spatial resolution showing sharp lines (D$_1$ and D$_2$) attributed to excitons bound to shallow impurities. (d) Energies of different exciton species in a DQW as a function of the electric field. The arrows show possible trion dissociation processes. 
}
\end{figure}

The optical studies were carried out at 4.2~K using a microscopic photoluminescence ($\mu$-PL) setup with spatial and spectral resolutions of 0.6~$\mu$m and 0.08~meV, respectively.  Figure~\ref{PLV}(a) displays the bias dependence of the PL intensity $I_{PL}$ recorded by exciting the sample with a focused  laser beam  (diameter $\phi_L=1.2~\mu$m) and integrating the emission over a 7.5~$\mu$m-long and 1~$\mu$m-wide slit across the Ti gate. The laser wavelength and power were $\lambda=780$~nm (1.59~eV) and $P=240$ nW, respectively. Further experimental details can be found in the supplementary material (SM), Sec.~SM1.%\ref{Spectroscopic_experiments}. 
The PL bias map shows the typical spectral lines from DXs, IXs, and Ts.\cite{Schinner_PRB83__11,PVS266} The strong trion signal signalizes the availability of free carriers. 
The energy and intensity of T line remain constant over a wide range of biases denoted as FB in Fig.~\ref{PLV}(a) (i.e., for biases $V_b$ between 0.5 and 1.2~V compensating the built-in potential of the Schottky junction). Within this range, the flow of free carriers between the QWs screens the applied electric field across the DQW structure, thus  inhibiting IX formation.
The transition region between the FB and IX regimes is characterized by an enhanced DX emission together with a reduction of the trion intensity. The latter is attributed to the depletion of free carriers in one of the QWs as these carriers tunnel to and accumulate in  the adjacent QW.\cite{Schinner_PRB83__11,Schinner_PRB87_41303_13} 
Carrier accumulation in one of the QWs also accounts for the non-linear bias dependence of the IX energy close to the FB-IX transition.\cite{Schinner_PRB83__11}

%%%%%%%%%%%%%%%%%%%%%%%%%%%%%%%
%%%%%%%%%%%%%%%%%%%%%%%%%%%%%%%

More interesting spectral features appear when mapping the FB-IX transition range using a much lower excitation densities, as illustrated in  Fig.~\ref{PLV}(b). Strikingly, the IX line does not smoothly ``branch off'' from  the DX and T lines but only appears  for $V_b<0.25$~V, thus indicating that its formation is blocked within the  bias range BR indicated in the figure. The latter results in a spectral gap between T energy and the onset of the IX emission given by $\Delta E_B=5.7$~meV. 
Furthermore, the IX appearance at $V_b\sim0.25$~V is accompanied by a drastic reduction of the T intensity, which indicates that the IXs result from the dissociation of T states. 
Finally, several sharp emission lines appear in the IX blockade regime BR. These lines can be grouped into two families (labeled as DX$^b$ and IX$^b$ and indicated by the dot-dashed lines in Fig.~\ref{PLV}(b))  with bias dependence similar to the ones for free DX and IX species, respectively. The energies of the individual lines are typically between 1.5 to 10~meV lower than the T emission.

The observation of multiple sharp lines in Fig.~\ref{PLV}(b) arises from the integration of the emission over an extended area on the sample surface. Figure~\ref{PLV}(c) displays a similar bias map recorded by collecting the emission over a much smaller sample area (corresponding to the resolution limit of $0.4~\mu$m$^2$), where individual lines can be observed (labeled as  D$_1$ and D$_2$). The energy distribution of such emission lines cover the typical energy range of excitons bound to shallow impurities. Natural candidates for the impurities are silicon (donor) and carbon (acceptor), which are the most common impurities present in the MBE growth.   Interestingly, these lines mainly appear over the narrow bias range  BR and are generally not present in the IX bias range, where their excitation should  also be energetically favorable. This behavior indicates that these individual states are  excited via a resonant process taking place over a narrow range of biases. 

The correlation between the appearance of bound exciton and IX lines, on the one hand, and the disappearance of T on the other hand, suggests that the former results from dissociation of T, governed by energy conservation. The electric field dependence of the energy $E_i$ of the different excitonic species ($i=$DX, IX, T)  is summarized in Fig.~\ref{PLV}(c).\footnote{we use electric field instead of bias to take into account screening effects.}  The non-monotonous electric field dependence of $E_{\rm IX} (F_z)$  results from the combined effect of the QCSE  and the electric field dependence of IX binding energy $\Delta E_{\rm IX}$.~\cite{Takahashi_JAP76_2299_94} Two trion dissociation scenarios involving resonant tunneling and  IX formation can be envisaged. For simplicity, we will consider electron trions  (a similar argument  applies to hole trions). In the first, a trion dissociates via the emission of a free carrier in the same QW (with energy equal to the band gap energy $E_\mathrm{g}$) and the tunneling of the extra electron to form an IX, as indicated by the solid curved arrows in Fig.~\ref{PLV}(d). Energy conservation  requires that
\begin{equation}
\label{eq1}
%E_{\rm IX}(F_z)-E_{\rm T} < -\Delta_1=\Delta E_\mathrm{T-DX}+\Delta E_{\rm DX},
E_{\rm T}-E_{\rm IX}(F_z) \geq E_{\rm g}-E_\mathrm{T}=\Delta_1.
%E_{\rm IX}(F_z)-E_{\rm T} < -\Delta_1=E_\mathrm{T}-E_{\rm g},
\end{equation} IX formation will thus be blocked within an energy range  $\Delta_1$ below the trion energy.
Alternatively,  the free electron of the previous process may form a DX in the presence of a free hole.  In this case, the condition to lift the IX formation blockade reads:
\begin{equation}
\label{eq2}
%E_{\rm IX}(F_z)-E_{\rm T} < -\Delta_2=\Delta E_\mathrm{T-DX}.
E_{\rm T}-E_{\rm IX}(F_z) \geq E_{\rm{DX}}-E_{\rm T}=\Delta_2.
%E_{\rm IX}(F_z)-E_{\rm T} < -\Delta_2=E_{\rm T} - E_{\rm{DX}}.
\end{equation}
 Since this process requires a an extra particle, it is expected to be suppressed at low excitation densities.

According to the calculations in  Ref.~\onlinecite{Takahashi_JAP76_2299_94}, the binding energy of DX, $\Delta E_{\rm DX}=E_\mathrm{DX}- E_\mathrm{g}\approx -5.7$~meV for the DQW structure used here. From Fig.~\ref{PLV}(a), we obtain $\Delta_2= 1.6$~meV, thus yielding $\Delta _{1}=7.3$~meV. The measured $\Delta E_B$ at different laser powers (cf. Fig.~\ref{PLV} and SM2) typically yield $\Delta E_B=5\pm 1$~meV, in reasonable agreement with $\Delta _{1}$ for low excitation densities. At high excitation power and high temperature, the blockade feature cannot be identified anymore (see~SM2).

The energy conservation constraint imposed by Eqs.~\ref{eq1} and \ref{eq2} stabilizes trions and opens a gap in the emission spectrum of free IX species.  Within the gap, trions can still be  resonantly converted into lower energy excitons bound to an impurity, e.g. DX$^b$ and IX$^b$ (their energies labeled as $E_{\rm DX^b}$ and $E_{\rm IX^b}$, respectively), indicated by the straight arrows in Fig.~\ref{PLV}(d). We illustrate two possible mechanisms for the conversion. In the first case, the T in one QW dissociates into a free particle and a DX bound to a shallow center in the same QW (DX$^b$). In the second case, one of the trion particles tunnels across the barrier to form an IX$^b$ bound to a shallow center in the adjacent QW.  (Note that the free particle can also be replaced by a DX if an extra hole is available.) These two processes yield bound species with the bias dependences corresponding to the ones for the DX$^b$ and IX$^b$ states\cite{Brum_PRB32_2378_85,Zhao_PRB45_11346_92} indicated by the dot-dashed lines in Fig.~\ref{PLV}(b) (e.g., both D$_1$ and D$_2$ in Fig. \ref{PLV}(c) behave as DX$^b$).

%%%%%%%%%%%%%%%%%%%%%%%%%%%%%%%
\begin{figure}
\includegraphics[width=0.9\columnwidth, angle=0, clip]{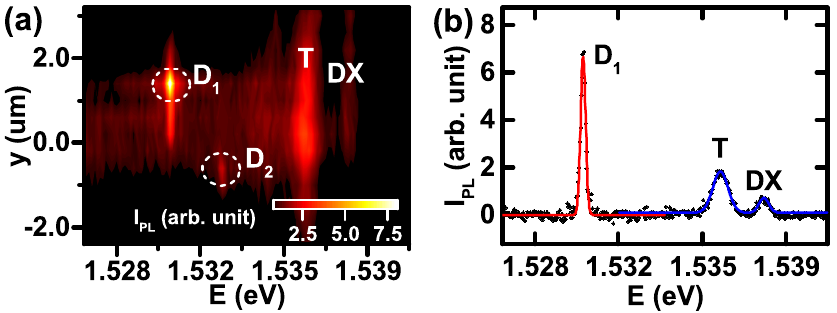}% Here is how to import EPS art
\caption{
\label{spa}
(a) Spectral PL image showing the spatial distribution of free excitons (DX and T) and excitons bound to shallow centers (D$_{1}$ and D$_{2}$). The excitons were remotelly excited by a focused laser spot placed at $y=0$.  The DQW is biased at the edge of the flatband. 
(b) PL spectrum (background subtracted) at the location of D$_{1}$ (symbols). The superimposed lines are Gaussian fits.
}
\end{figure}
%%%%%%%%%%%%%%%%%%%%%%%%%%%%%%%
%1529.4~meV  (810.7~nm)1531.6~meV (809.5~nm)

We now concentrate on the properties of the bound exciton states. Their spatial distribution was determined by spatially resolved PL maps recorded while biasing the structure in the BR, as illustrated in Fig.~\ref{spa}(a). The sample was excited by a focused laser spot at $y=0$ and the emission recorded as a function of the distance from the excitation spot.  In addition to the lines from DX and T, the figure shows localized peaks from two defect centers D$_{\rm 1}$ and D$_{\rm 2}$ at 1530.4~meV  and 1532.7~meV, respectively (the same defects as in Fig.~\ref{PLV}(c)), which are populated by carriers diffusing from the excitation spot. The shallow centers are typically between 2 to 4~$\mu$m apart, thus yielding an areal density  $n_d\approx 10^7$cm$^{-2}$. Figure~\ref{spa}(b) displays a cross-section of Fig.~\ref{spa}(a) at the position of D$_{\rm 1}$ (maximum emission) at  $y=1.4~\mu$m. 
The spectral line shape is fitted by a  Gaussian profile (red) with a full width at half maximum of $\hbar\Gamma_S=0.26$~meV. The latter is significantly narrower than the DX and T linewidths (of 0.42~meV and 0.66~meV, respectively, cf. blue curves).
The temperature dependence of the emission intensity of the bound exciton (see SM3 for details) yields an activation energy $E_A$ comparable to the energy red-shift with respect to T, thus indicating that the decay is caused by thermal dissociation into T. 
The PL intensity and linewidth, however, saturate for temperature lower than 5~K. The line width broadening at 4~K is probably due to charge fluctuation.\cite{Bayer_PRB65_041308_02}

%--------------------------------------------
%\section{Magnetic field dependence}
%%%%%%%%%%%%%%%%%%%%%%%%%%%%%%%
\begin{figure}
\includegraphics[width=0.9\columnwidth, angle=0, clip]{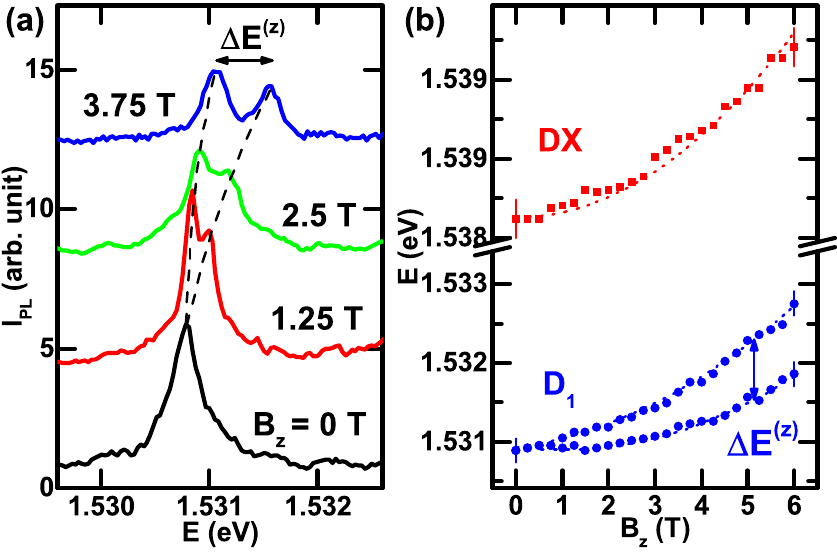}% Here is how to import EPS art
\caption{
\label{Bdep} 
(a) PL spectra of D$_1$ recorded under different magnetic fields $B_z$ applied along the growth axis (the curves are shifted vertically for clarity)  showing the Zeeman splitting $\Delta E_z$ as well as the diamagnetic blue-shift $\Delta E^{(d)}$.  (b) Peak energies of the DX (squares) and D$_{1}$ (circles) resonances as a function of $B_z$. The dashed lines are fits to Eq.~\ref{EqEB}. Due to the larger linewidths, the Zeeman splitting cannot be clearly resolved for the DXs. 
}
\end{figure}
%
%%%%%%%%%%%%%%%%%%%%%%%%%%%%%%%
Information about the internal structure of the bound excitons was obtained from measurements under a magnetic field $B_z$ applied along the DQW growth axis (cf. Fig.~\ref{Bdep}(a)). The field splits the emission peak into a Zeeman doublet with energy splittings $\Delta E^{(z)}$ and average energies increasing  with $B_z$. The $B_z$ dependence of the doublet energy is summarized by the blue circles in Fig.~\ref{Bdep}(b) and can be expressed as:
\begin{equation}
E=\Delta E^{(d)}\pm \frac{1}{2}\Delta E^{(z)} = \frac{e^2}{8\mu}\langle\rho^2\rangle B_z^2 \pm\frac{1}{2} g \mu_B B_z.
\label{EqEB}
\end{equation}
 Here,  $e$ is the electron charge, $\mu$ the in-plane reduced excitonic mass satisfying $\frac{1}{\mu}=\frac{1}{m_e}+\frac{1}{m_{hh}}$, where $m_e=0.0665~ m_0$ and $m_{hh}=0.34~m_0$ denote the in-plane electron and heavy-hole masses, respectively, and  $m_0$ is the electron rest mass.\cite{Molenkamp_PRB38_4314_88} 
$\mu_B$ is the Bohr magneton and $g$ the exciton Land{\' e} $g$-factor, which depends on the corresponding $g$-factors for holes and electrons.\cite{Snelling92a} The first term on the r.h.s is  the diamagnetic blue-shift $\Delta E^{(d)}$ resulting from the confinement of excitonic wave function by $B_z$, which depends on the spatial extent of the exciton wave function $\rho$.\cite{Nash_PRB_39_10943_89,Walck_PRB_57_9088_98}

The blue dashed lines superimposed on the experimental points are fits to Eq.~\ref{EqEB}, which  yield $g=2.2$ and $\rho=9$~nm. This value for the effective exciton radius is in good agreement with the one reported for excitons bound to shallow donors.\cite{Brozak_PRB_40_1265_89}. Figure~\ref{Bdep}(b) also shows the magnetic field dependence of the DX resonance (red squares). Due to the large spectral linewidth, the Zeeman splitting cannot be clearly resolved. From the quadratic dependence of the diamagnetic shifts (cf. Eq.~\ref{EqEB}), we extract an exciton radius of 12~nm. In addition to the quadratic diamagnetic behavior, one also observes small discontinuities in the field dependence, probably due to Landau quantization of the free carriers.

%%%%%%%%%%%%%%%%%%%%%%%%%%%%%%%
\begin{figure}
\includegraphics[width=0.9\columnwidth, angle=0, clip]{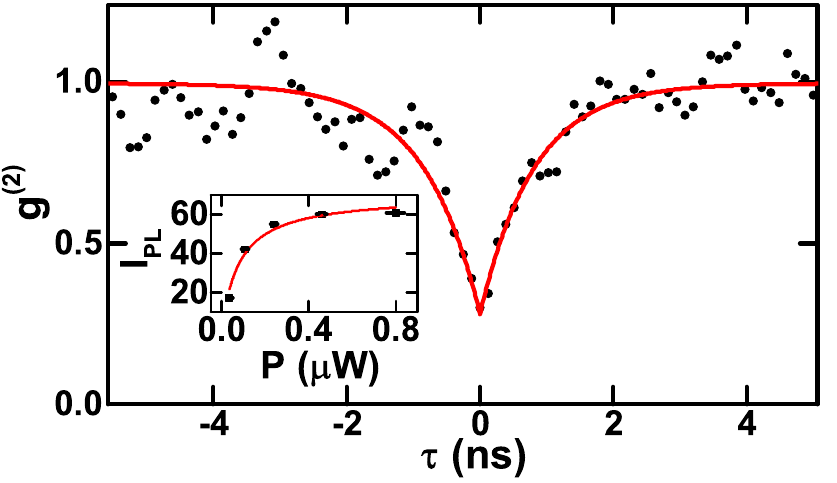}% Here is how to import EPS art
\caption{\label{sp} Photon auto-correlation $g^{(2)}$ for center D$_1$ in Fig.~\ref{PLV}(c).
The excitation spot was offset from the detection spot to reduce the background emission. The contribution from the background has been corrected. Single-photon nature is confirmed by the observation of $g^{(2)}(0)<0.5$. Inset: power dependence of emission intensity (symbols). The superimposed line is a fit to Eq.~\ref{P} yielding $P_0=77$~nW.
}
\end{figure}
%%%%%%%%%%%%%%%%%%%%%%%%%%%%%%%

Motivated by the narrow linewidths of the bound excitons as well as their spectral separation from free exciton states, we investigate the bound-excitons' potential as two-level systems emitting  single photons. The two-level characteristic can be inferred by the saturation of emission intensity with increasing laser excitation powers, as indicated by the symbols in Fig.~\ref{sp} inset. The solid line is a fit to a model for light scattered by two-level systems\cite{Mollow_PR_188_1969}:

\begin{equation}\label{P}
I_{PL}=\frac{I_0 P}{P_0+P}.
\end{equation}

\noindent Here, $I_0$ is the intensity limit approached at high input power and $P_0$ is the saturation input power, defined as the input power at which the intensity reaches $\frac{1}{2}I_0$. 
The saturation points to a finite number of two-level centers and also  hinders the observation of bound excitons under high excitation power, as observed in Fig.~\ref{PLV}(a).
%The saturation of the emission intensity provides a strong evidence for single-photon character of the emission, since a two-level system can only be re-excited when in the ground state. 

In order to confirm the single-photon nature of the emission, we measured the second-order auto-correlation function $g^{(2)}(\tau)$ of the center D$_1$ of Fig.~\ref{PLV}(c) using a Hanbury-Brown and Twiss setup. 
The $g^{(2)}$ histogram in Fig.~\ref{sp} shows the characteristic dip at zero delay $\tau=0$, signalizing photon anti-bunching.  
The solid line is a fit to delay dependence of the form $g^{(2)}(\tau)=1-(1/N_p) \exp (-|\tau|/\tau_R)$. The fit yields an average  number of simultaneously emitted photons at zero delay  $N_p<0.5$,  thus confirming the single-photon nature of the emission. Finally, the fitted decay time constant $\tau_R=0.8$~ns  is  comparable with the values measured for self-assembled  InAs/GaAs QDs.\cite{Zwiller01a}

In conclusion, we have reported the selective excitation of  single excitons bound to shallow impurity centers via resonant tunneling in GaAs DQW structures. The observation of these centers becomes possible because of the blockade of carrier tunneling to form intrinsic exciton states over a range of applied electric fields. The latter hinders the formation of IXs, thus opening a gap in the emission spectrum. The PL within this gap is dominated by the dissociation of trions via tunneling process involving excitons bound to isolated shallow impurity centers.  These bound excitons behave as two-level systems emitting single photons, as confirmed by the observation of  anti-bunching in the photon autocorrelation histogram. 

%The narrow distribution of emission energies as compared to self-assembled quantum dots makes the bound exciton emitters potential candidates for single-photon sources.%, with  important implications for both basic and applied research.   
Single-photon emitters based on shallow centers profit from the well-developed semiconductor processing techniques. 
A major advantage of these  centers is the much narrower distribution of emission energies as compared, for instance, to single-photon sources based on self-assembled quantum dots.
In contrast to previously reported impurity-based single-photon emitters on (Al,Ga)As structures,\cite{minari2012,Ikezawa2012} these bound excitons  are electrically controlled and can be easily integrated with other  functionalities such as site control and spin injection. 
The electrical control enables the isolation of individual centers by in-situ fabrication of electrostatic gates.\cite{Gschrey_APL102_251113_13} 
The shallow centers in GaAs DQWs are promising single photon emitters with both optical and electrical controls.

\begin{acknowledgments}
 We thank S. Ludwig  and M. Ramsteiner for many helpful discussions and suggestions. We thank S. Takada, C. B\"auerle, S. Rauwerdink, S. Meister and C. Hubert for support in sample fabrication. We acknowledge financial support from the German Forschungsgemeinschaft, DFG.
\end{acknowledgments}

%\bibliography{mypapers,literature}

%

%%%%%%%%%%%%%%%%%%%%%%%%%%%%%%%%%%%%%%%%%%
\clearpage

%\SupplementaryMaterials 

%%%%%%%%%%%%%%%%%%%%%%%%%%%%%%%%%%%%%%%%%%
% commands to change labeling in the Suppl. Material Section
\renewcommand{\thesection}{SM\arabic{section}} 
\setcounter{section}{0}  
\renewcommand{\thefigure}{SM\arabic{figure}} 
\setcounter{figure}{0}  
\renewcommand{\theequation}{SM\arabic{equation}} 
\setcounter{equation}{0}  
%%%%%%%%%%%%%%%%%%%%%%%%%%%%%%%%%%%%%%%%%%

\begin{widetext}
\section*{Supplementary Material  for:\\
Tunneling blockade and single-photon emission in GaAs double quantum well structures}
\end{widetext}

\section{Spectroscopic experiments}
\label{Spectroscopic_experiments}

The microscopic photoluminescence ($\mu$-PL) experiments were carried out in a helium bath cryostat at 4.2 K (Attocube Confocal Microscope) with positioning control provided by a piezoelectric stage. The incoming laser with a wavelength of 780~nm  was focused on the sample surface using an objective with numerical aperture of NA$\sim$0.8. The laser energy lies below the band-gap of the Al$_{0.3}$Ga$_{0.7}$As barriers and thus selectively excites electron hole pairs only in the GaAs QWs. The diameter of the laser spot $\phi_L=1.2~\mu$m sets the spatial resolution to $0.6~\mu$m.
The PL was collected by the same objective and coupled to a monochromator using either a single-mode optical fiber (for confocal measurements) or by a fiber bundle (for spatially resolved measurements). The input slit of the monochromator was selected to yield a spectral resolution of 0.08~meV.  A liquid-nitrogen cooled CCD detector was used to detect the light at the output port of the monochromator. The dimensions of each CCD pixel corresponds to 0.35~$\mu$m on the sample.

The photon auto-correlation studies  were carried out using a Hanbury-Brown and Twiss setup. The PL from the shallow center was collected by the objective, filtered by a Semrock band-pass filter with a bandwidth of 810$\pm$1.5~nm
%{\bf (PVS: is that really only 1.5 nm?)(MY: yes. bandwidth 3.1 nm}, 
and then coupled to a single-mode fiber connected to a 50/50 fiber splitter. The outputs of the fiber splitters were then  sent to a pair of superconducting  single-photon detectors (Single Quantum Eos). The coincidence statistics was  performed by a PicoQuant Picoharp photon counting system with a time bin of 128~ps. 

% ---------------------------------------------------
\section{Blockade feature under varying power and temperature}
\label{Blockade}

%%%%%%%%%%%%%%%%%%%%%%%%%%%%%%%%%%%%%%%%%
\begin{figure}
\includegraphics[]{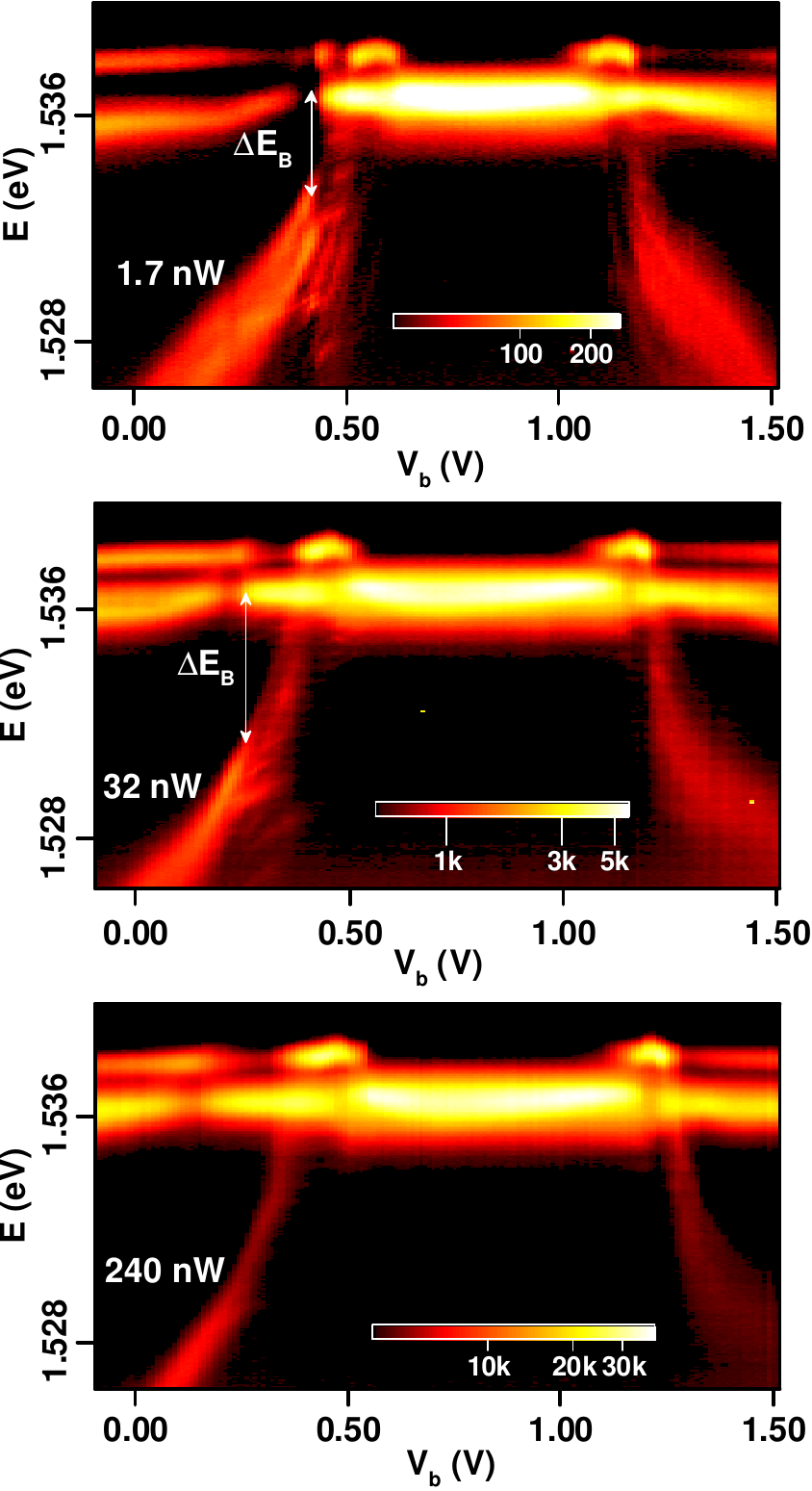}
\caption{\label{varP}Spatially integrated PL as a function of bias voltage $V_b$ recorded under different optical excitation power}
\end{figure}
%%%%%%%%%%%%%%%%%%%%%%%%%%%%%%%%%%%%%%%%%

%%%%%%%%%%%%%%%%%%%%%%%%%%%%%%%%%%%%%%%%%
\begin{figure}
\includegraphics[]{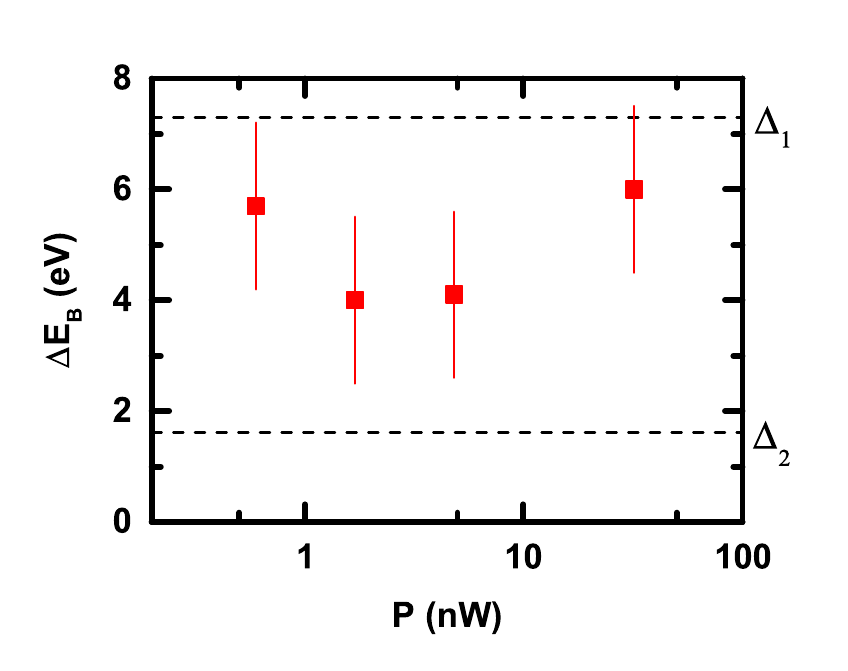}
\caption{\label{Eb}Blockade range $\Delta E_B$ extracted at different laser power $P$.}
\end{figure}
%%%%%%%%%%%%%%%%%%%%%%%%%%%%%%%%%%%%%%%%%

%%%%%%%%%%%%%%%%%%%%%%%%%%%%%%%%%%%%%%%%%
\begin{figure}
\includegraphics[]{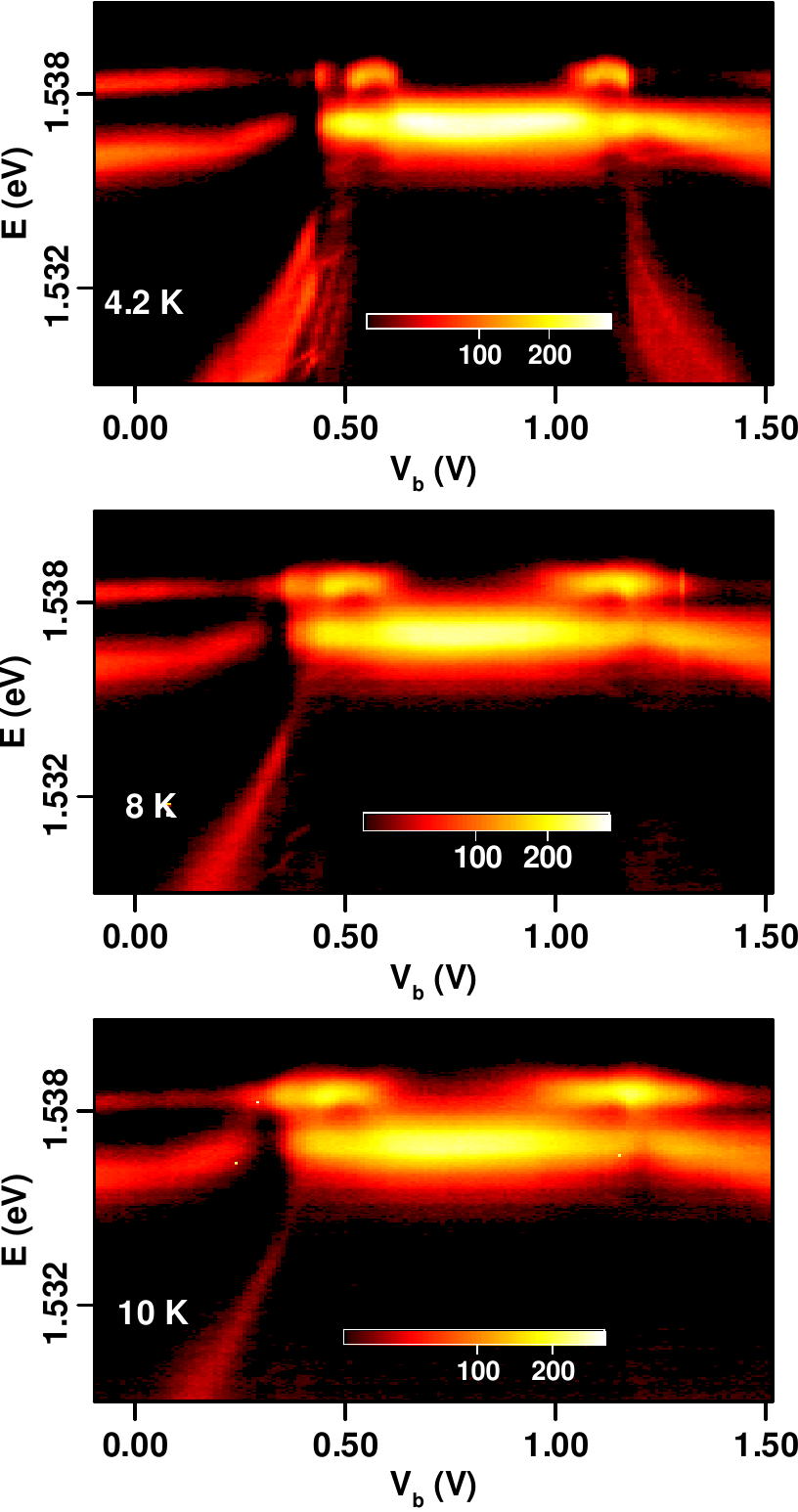}
\caption{\label{varT}Spatially integrated PL as a function of bias voltage $V_b$ recorded under different temperature. Laser power is kept at 1.7~nW.}
\end{figure}
%%%%%%%%%%%%%%%%%%%%%%%%%%%%%%%%%%%%%%%%%

Spatially integrated PL vs. bias voltage $V_b$ maps were measured at 4.2~K for different laser excitation powers. A few examples are shown in Fig.~\ref{varP}. At high excitation power the blockade is no longer visible anymore. The blockade energy $\Delta E_B$ is extracted as the energy difference between T and the onset energy of the IX detected when the bias is tuned beyond the flat-band range. We plot $\Delta E_B$ for the lowest four excitation power $P$ in Fig.~\ref{Eb} and find that the blockade stays between $\Delta _{1}$ and $\Delta _{2}$ defined by Eqs.~1 and 2 of the main text.

As expected, rising temperature smears out the blockade feature, making it harder to identify, as shown in Fig.~\ref{varT}.

% ---------------------------------------------------
\section{Activation energy}
\label{Activation_energy}

\begin{figure}
\includegraphics{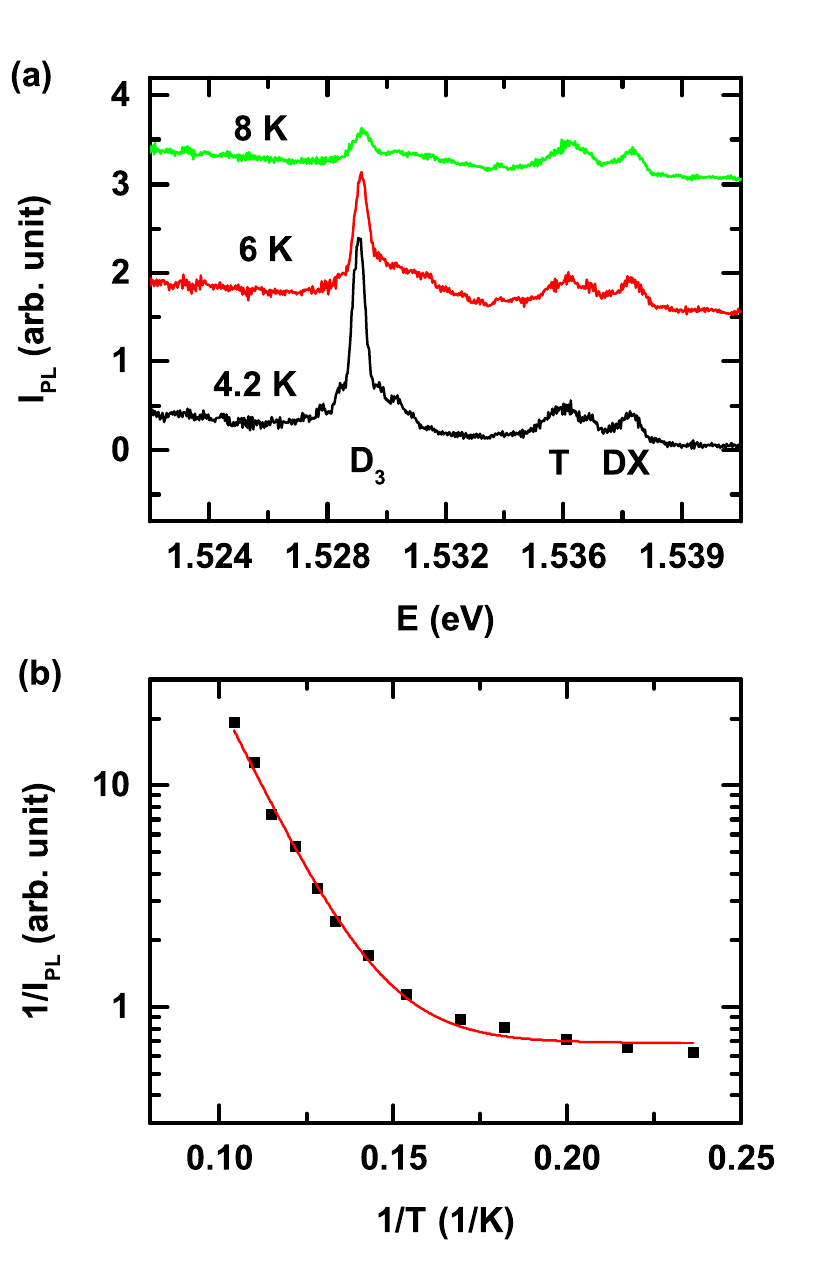}
\caption{\label{Arh} (a) Spectrum taken at different temperature, showing the thermal quenching of $I_{PL}$ from D$_3$ at 1.529~eV. (b) The inverse of PL in logarithmic scale vs. the inverse of temperature. Solid: data. Line: fit.}
\end{figure}

By varying the sample temperature $T$ we thermally quench the PL intensity for another bound exciton, D$_3$ at 1.529~eV, from which we extract the activation energy $E_A$ for its dissociation. The probability of thermal activation obeys the Boltzmann's law

\begin{equation}
\label{Boltz}
p\propto{\rm{exp}}\left[-\frac{E_A}{k_BT}\right].
\end{equation}
Here, $k_B$ is the Boltzmann constant. The PL quantum efficiency can be expressed as
\begin{equation}
\eta=\frac{1}{1+\xi{\rm{exp}}(-E_A/k_BT)},
\end{equation}

\noindent where $\xi$ is a constant. The intensity of PL is proportional to its quantum efficiency, and thus decreases with rising temperature, as can be seen in Fig.~\ref{Arh}(a).

We can express the inverse of the PL intensity, letting $\alpha=1/T$,  as
\begin{equation}\label{Tfit}
\frac{1}{I_{PL}}=A+B~{\rm{exp}}\left(-\frac{E_A}{k_B}~\alpha\right),
\end{equation}
where $A$ and $B$ are additional constants. By fitting the data with Eq.~\ref{Tfit} (see Fig.~\ref{Arh}(b)), we extract $E_A=6.5$~meV, which is comparable with the energy difference between trion T and the defect center D$_3$, 7~meV.

%%%%%%%%%%%%%%%%%%%%%%%%%%%%%%%%%%%%%%%%%
%%%%%%%%%%%%%%%%%%%%%%%%%%%%%%%%%%%%%%%%%

% ---------------------------------------------------

\section{List of abbreviations and symbols}
\label{abbre}
\begin{tabular}{l l}
QW & quantum well\\
DQW & double quantum well\\
DX & direct exciton\\
T & trion\\
IX & indirect exciton\\
CB & conduction band\\
VB & valence band\\
QCSE~~ & quantum-confined stark effect\\
MBE & molecular beam epitaxy\\
PL & photoluminescence\\
$\mu$-PL & microscopic photoluminescence\\
SM & supplementary material\\
FB & flat-band\\
BR & blockade regime\\
DX$^{b}$ & bound exciton with DX-like bias-dependence\\
IX$^{b}$ & bound exciton with IX-like bias-dependence\\
D$_{1,2,3}$ & individual bound excitons\\
\end{tabular}
\\

\begin{tabular}{l l}
$F_z$ & transverse electric field\\
$V_b$ & bias voltage\\
$d_{QW}$ & QW thickness\\
$d_b$ & barrier thickness\\
$\phi_c$ & Ti contact diameter\\
$I_{PL}$ & PL intensity\\
$\phi_L$ & laser spot diameter\\
$\lambda$ & laser wavelength\\
$P$ & laser power\\
$\Delta E_B$ & energy gap between onset of IX and T\\
$E_{\rm IX}$ & IX energy\\
$E_{\rm T}$ &T energy\\
$E_{\rm DX}$ &DX energy\\
$E_g$ & bandgap energy\\
$\Delta E_{\rm IX}$ & IX binding energy relative to bandgap\\
$\Delta E_{\rm T-DX}$ & T binding energy relative to DX\\
$\Delta E_{\rm DX}$ & DX binding energy relative to bandgap\\
$\Delta_1$ & energy difference between T and bandgap\\
$\Delta_2$ & energy difference between T and DX\\
$n_d$ & areal density of shallow impurities\\
$\Gamma_S$ & spectral linewidth in frequency\\
$\Delta E_A$ & activation energy\\
$\Delta E^{(d)}$ & diamagnetic energy shift\\
$\Delta E^{(z)}$ & Zeeman splitting\\
$e$ & electron charge\\
$\mu$ & reduced excitonic mass\\
$m_e$ & electron mass\\
$m_{hh}$ & heavy hole mass\\
$\rho$ & extent of exciton wave function\\
$B_z$ & transverse magnetic field\\
$g$ & Land{\' e} g-factor\\
$\mu_B$ & Bohr magneton\\
$g^{(2)}$ & second-order auto-correlation function\\
$\tau$ & time delay\\
$N_p$ & number of simultaneously emitted photons\\
$\tau_R$ & anti-bunching time constant\\
$\Gamma_R$ & anti-bunching linewidth in frequency\\
\end{tabular}

\end{document}